\documentclass[11pt,pre,floats,aps,nofootinbib]{revtex4}

\usepackage{amsmath,amssymb}
\usepackage{graphicx}

\def\ux{\underline{x}}
\def\uy{\underline{y}}
\def\uX{\underline{X}}
\def\uY{\underline{Y}}
\def\u0t{\underline{\tt 0}}
\def\0t{{\tt 0}}
\def\1t{{\tt 1}}
\def\cC{{\cal C}}
\def\cX{{\cal X}}
\def\cY{{\cal Y}}
\def\Rho{P}
\def\ldpc{{\rm LDPC}}
\def\G{{\sf G}}
\def\ed{\stackrel{{\rm d}}{=}}
\def\E{{\mathbb E}}
\def\ind{{\mathbb I}}
\def\B{{\sf B}}
\def\mh{\hat{\mu}}
\def\argmax{{\rm argmax}}

\def\BSC{{\rm BSC}}
\def\vh{\hat{v}}
\def\atanh{{\rm atanh}}
\def\?{{\tt ?}}

\def\sBP{{\tiny\rm BP}}
\def\de{{\rm d}}

\begin{document}

\title{Two Lectures on Iterative Coding and Statistical Mechanics}

\author{Andrea Montanari}

\affiliation{Laboratoire de Physique Th\'{e}orique de l'Ecole Normale
Sup\'{e}rieure\footnote {UMR 8549, Unit{\'e} Mixte de Recherche du
Centre National de la Recherche Scientifique et de l' Ecole Normale
Sup{\'e}rieure. }, 24, rue Lhomond, 75231 Paris CEDEX 05, France}

\date{\today}

\begin{abstract}
These are the notes for two lectures delivered at the Les Houches
summer school {\it Mathematical Statistical Mechanics}, held in 
July 2005. 

I review some basic notions on sparse graph error
correcting codes with emphasis on `modern' aspects,
such as, iterative belief propagation decoding.
Relations with statistical mechanics, inference and random combinatorial
optimization are stressed, as well as some general mathematical ideas 
and open problems.
\end{abstract}

\maketitle

%
%
\section{Introduction}

Imagine to enter the auditorium and read the following (partially erased)
phrase on the blackboard\\
\vspace{0.2cm}
{\sf TH*\phantom{AA}  L*CTU*E\phantom{AA} OF\phantom{AA} *********** 
\phantom{AA}WA*\phantom{AA} EX**EMELY\phantom{AA} B*RING}.\\ 
\vspace{0.2cm}
You will be probably able to reconstruct most of the 
{\it words} in the phrase despite the erasures. The reason is
that English language is redundant. One can roughly quantify 
this redundancy as follows. The English dictionary contains about 
$10^6$ words including technical and scientific terms. On the other 
hand, the average length of these words is about 
$8.8$ 
letters~\footnote{This estimate was obtained by the author averaging over 
$40$ words randomly generated by the site
{\tt http://www.wordbrowser.net/wb/wb.html}.}.
A conservative estimate of the number of `potential' English words
is therefore $26^8\approx 2\cdot 10^{11}$. A tiny fraction 
(about $10^{-5}$) of these possibilities is realized~\cite{English}. This is 
of course a waste from the point of view of information storage, but
it allows for words to be robust against errors (such as the above 
erasures). Of course, they are not infinitely robust: above some 
threshold the information is completely blurred out by noise
(as in the case of the name of the speaker in our example).

\begin{figure}
\begin{center}
\includegraphics[width=7cm]{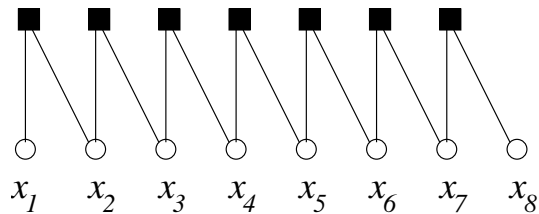}
\end{center}
\caption{A (too) na\"ive model for the English language.}
\label{fig:EightLetters}
\end{figure}
A very na\"ive model for the redundancy of English could be
the following. In order for a word to be easily pronounced, it must 
contain some alternation of vowels and consonants. Let us be rough
and establish that an English word is a sequence of $8$ letters, 
not containing two consecutive vowels and consonants.
This yields $2\cdot 21^4 5^4\approx 2.4\cdot 10^8$ distinct words,
which overestimates the correct number `only' by a factor $240$.
A graphical representation of this model is reproduced
in Fig.~\ref{fig:EightLetters}.

The aim of coding theory is to construct an optimal `artificial 
dictionary' allowing for reliable communication through unreliable 
media. It is worth introducing some jargon of this discipline.
Words of natural languages correspond to {\bf codewords} in coding.
Their length (which is often considered as fixed) is called the 
{\bf blocklength}: we shall denote it by $N$ throughout
these lectures. The dictionary (i.e. the set of all words used 
for communication) is called {\bf codebook} and denoted as 
${\cal C}$. As in our example, the dictionary size is usually 
exponential in the blocklength $|{\cal C}| = 2^{NR}$ and $R$
is called the code {\bf rate}. 
Finally, the communication medium referred to as the {\bf channel} and 
is usually modeled in a probabilistic way 
(we shall see below a couple of examples).
%
%
\section{Codes on graphs}

We shall now construct a family of such `artificial dictionaries'
(codes). For the sake of simplicity, codewords will be formed over the
binary alphabet $\{\0t,\1t\}$. Therefore a codeword 
$\ux\in\cC$ will be an element of the Hamming space
$\{\0t,\1t\}^N$ or, equivalently a vector of coordinates
$(x_1,x_2,\dots,x_N)\equiv \ux$. 

\begin{figure}
\begin{center}
\includegraphics[width=7cm]{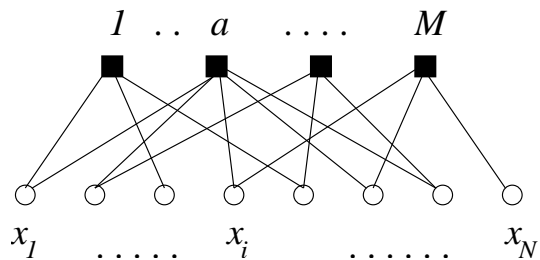}
\end{center}
\caption{Example of a Tanner graph.}
\label{fig:Tanner}
\end{figure}
The codebook $\cC$ is a subset of $\{\0t,\1t\}^N$. Inspired by our
simple model of English, we shall define $\cC$ by stipulating 
that $\ux$ is a codeword if and only if a certain number $M$
of constraints on the bits $x_1,\dots,x_N$ are met.
In order to specify these constraints, we will draw a 
bipartite graph ({\bf Tanner graph}) over vertices sets $[N]$ and $[M]$.
Vertices in these sets will be called, respectively, 
variable nodes (denoted as $i,j,\dots$) and check nodes ($a,b,\dots$).
If we denote by $i^a_1,i^a_2,\dots,i^a_k$ the variable nodes 
adjacent to check node $a$ in the graph, then 
$x_{i^a_1},x_{i^a_2},\dots,x_{i^a_k}$ must satisfy some constraint
in order for $\ux$ to be a codeword. An example of a Tanner graph 
is depicted in Fig.~\ref{fig:Tanner}.

Which type of constraints are we going to enforce on the
symbols $x_{i^a_1},x_{i^a_2},\dots,x_{i^a_k}$ adjacent to the same check?
The simplest and most widespread choice is a simple parity check
condition: $x_{i^a_1}\oplus x_{i^a_2}\oplus\cdots\oplus x_{i^a_k} =\0t$
(where $\oplus$ is my notation for sum modulo $2$). We will stick
to this choice, although several of the ideas presented below are
easily generalized. Notice that, since the parity check constraint is linear 
in $\ux$, the code $\cC$ is a linear subspace of $\{\0t,\1t\}^N$,
of size $|\cC|\ge 2^{N-M}$ (and in fact $|\cC|=2^{N-M}$ unless redundant 
constraints are used in the code definition). For general information
theoretic reasons one is particularly interested in the limit of large 
blocklength $N\to\infty$ at fixed rate. This implies that the number of 
checks per variable is kept fixed: $M/N = 1-R$.

Once the general code structure is specified, it is useful to
define a set of parameters which characterize the code. Eventually, 
these parameters can be optimized over to obtain better 
error correction performances. A simple set of such parameters
is the {\bf degree profile} $(\Lambda,\Rho)$ of the code. 
Here $\Lambda = (\Lambda_0,\dots,\Lambda_{l_{\rm max}})$, where
$\Lambda_l$ is the fraction of variable nodes of degree $l$
in the Tanner graph.
Analogously, $\Rho = (\Rho_0,\dots,\Rho_{k_{\rm max}})$, where
$\Rho_k$ is the fraction of check nodes of degree $k$.

Given the degree profile, there is of course a large number of graphs
having the same profile. How should one chose among them?
In his seminal 1948 paper, Shannon first introduced the idea
of randomly constructed codes. We shall follow his intuition 
here and assume that the Tanner graph defining $\cC$ is generated
uniformly at random among the ones with degree profile
$(\Lambda,\Rho)$ and blocklength $N$. The corresponding
code (graph) ensemble is denoted as $\ldpc_N(\Lambda,\Rho)$
(respectively $\G_N(\Lambda,\Rho)$), an acronym for 
{\bf low-density parity-check} codes. 
Generically, one can prove that
some measure of the code performances concentrates in probability
with respect to the choice of the code in the ensemble.
Therefore, a random code is likely to be (almost) as good as
(almost) any other one in the ensemble. 

An particular property of the random graph ensemble will be 
useful in the following. Let $G\ed \G_N(\Lambda,\Rho)$ and
$i$ a uniformly random variable node in $G$. Then, with high 
probability (i.e. with probability approaching one in the large
blocklength limit), the shortest loop in the $G$ through 
$i$ is of length $\Theta(\log N)$.
%
%
\section{A simple--minded bound and belief propagation}

\subsection{Characterizing the code performances}

Once the code is constructed, we have to convince ourselves
(or somebody else) that it is going to perform well.
The first step is therefore to produce a model for 
the communication process, i.e. a noisy channel. A simple such model 
(usually referred to as {\bf binary symmetric channel}, or 
$\BSC(p)$) consists
in saying that each  bit $x_i$ is flipped independently of the
others with probability $p$. In other words the channel output
$y_i$ is equal to $x_i$ with probability $1-p$ and different with
probability $p$.
This description can be encoded in a transition probability 
kernel $Q(y|x)$. For $\BSC(p)$ we have
$Q(\0t|\0t) = Q(\1t|\1t) = 1-p$ and $Q(\1t|\0t) = Q(\0t|\1t) = p$.
More generally, we shall consider transition probabilities 
satisfying the `symmetry condition' $Q(y|\0t)= Q(-y|\1t)$
(in the $\BSC$ case, this condition fulfilled if we use the $+1$, $-1$
notation for the channel output).

The next step consists in establishing a measure of the performances
of our code. There are several natural such measures, for instance 
the expected number of incorrect symbols, or of incorrect words.
To simplify the arguments below, it is convenient to consider a slightly 
less natural measure, which conveys essentially the same information.
Recall that, given a discrete random variable $X$, with distribution 
$\{p(x): x\in \cX\}$, its entropy, defined as
\begin{eqnarray}
H(X) = -\sum_{x\in\cX}p(x)\log p(x)\, 
\end{eqnarray}
is a measure of how `uncertain' is $X$. Analogously, if $X,Y$
are two random variables with joint distribution 
$\{p(x,y): x\in \cX,\, y\in\cY\}$, the conditional entropy
\begin{eqnarray}
H(X|Y) = -\sum_{x\in\cX,\,y\in\cY}p(x,y)\log p(x|y)\, 
\end{eqnarray}
is a measure of how `uncertain' is $X$ once $Y$ is given.

Now consider a uniformly random codeword $\uX$ and the corresponding
channel output $\uY$ (as produced by the binary symmetric channel).
The conditional entropy $H(\uX|\uY)$ measures how many additional
bits of information (beyond the channel output) do we need for 
reconstructing $\ux$ from $\uy$. This is a fundamental quantity but
sometimes difficult to evaluate because of its non-local nature.
We shall therefore also consider the bitwise conditional entropy
\begin{eqnarray}
h_{{\rm b}} = \frac{1}{N}\sum_{i=1}^N H(X_i|\uY)\, .
\end{eqnarray}
%
%
%
\subsection{Bounding the conditional entropy}

Before trying to estimate these quantities, it is convenient to 
use the channel and code symmetry in order to simplify the task.
Consider for instance the conditional entropy. Denoting by
$\u0t$ the `all zero' codeword,  we have
\begin{eqnarray}
H(\uX|\uY) &= &-\sum_{\ux,\uy}p(\ux)p(\uy|\ux)\log p(\ux|\uy) =\\
&=& -\sum_{\uy}p(\uy|\u0t)\log p(\u0t|\uy) = \\
&=& -\E_y\log p(\u0t|\uy) \, ,\label{eq:Entropy}
\end{eqnarray}
where $\E_y$ denotes expectation with respect to the probability measure
$p(\uy|\u0t) = \prod_i Q(y_i|\0t)$. In the $\BSC(p)$ case, under this measure,
the $y_i$ are i.i.d. Bernoulli random variables with parameter $p$.
Furthermore, by using Bayes theorem, we get 
\begin{eqnarray}
H(\uX|\uY) &= & -\E_y\log p(\uy|\u0t)+ \E_y
\log\left\{\sum_{\ux\in \cC} p(\uy|\ux)\right\} =\\
& = & -N\sum_y Q(y|\0t)\log  Q(y|\0t) +
\E_y\log\left\{\sum_{\ux} \prod_iQ(y_i|x_i)\prod_a\ind(x_{i^a_1}\oplus
\cdots\oplus x_{i^a_k}=0) \right\}\, .
\end{eqnarray}
The first term is easy to evaluate consisting of a finite sum
(or, at most, a finite-dimensional integral). 
The second one can be identified as the quenched free energy 
for a disordered model
with binary variables (Ising spins) associated to vertices of the
the Tanner graph $G$.
Proceeding as in (\ref{eq:Entropy}) one also gets the following 
expression for the single bit conditional entropy
\begin{eqnarray}
H(X_i|\uY) =  -\E_y\log p(x_i=\0t|\uy)\, .
\end{eqnarray}

A simple idea for bounding a conditional entropy is to use the
`data processing inequality'. This says that, if 
$X\to Y\to Z$ is a Markov chain, then $H(X|Y)\le H(X|Z)$.
Let $\B(i,r)$ denote the subgraph of $G$ whose variable nodes lie 
at a distance at most $r$ from $i$ (with the convention that a 
check node $a$ belongs to $\B(i,r)$ only if all of the adjacent variable 
nodes belong to $G$).
Denote by $\uY_{i,r}$ the vector of output symbols $Y_j$, such that 
$j\in\B(i,r)$. The data processing inequality implies
\begin{eqnarray}
H(X_i|\uY) \le H(X_i|\uY_{i,r}) = -\E_y\log p(x_i =\0t|\uy_{i,r})\, .
\label{eq:UpperBoundEXIT}
\end{eqnarray}
A little more work shows that this inequality still holds if 
$p(x_i |\uy_{i,r})$ is computed as if there weren't parity checks 
outside $\B(i,r)$. In formulae, we can substitute 
$p(x_i = \0t |\uy_{i,r})$ with 
\begin{eqnarray}
p_{i,r}(x_i = x |\uy_{i,r}) \equiv \sum_{\ux_{i,r} = x}
p_{i,r}(\ux_{i,r} |\uy_{i,r})\, .
\end{eqnarray}
where
\begin{eqnarray}
p_{i,r}(\ux_{i,r} |\uy_{i,r}) \equiv \frac{1}{Z_{i,r}(\uy_{i,r})}\;
\prod_{j\in \B(i,r)}Q(y_j|x_j)\,
\prod_{a\in \B(i,r)}\ind(x_{i^a_1}\oplus\cdots\oplus x_{i^a_k}=0)\, .
\label{eq:TreeDistr}
\end{eqnarray}
and $Z_{i,r}(\uy_{i,r})$ ensures the correct normalization of
$p_{i,r}(\ux_{i,r} |\uy_{i,r})$.

\begin{figure}
\begin{center}
\includegraphics[width=5cm]{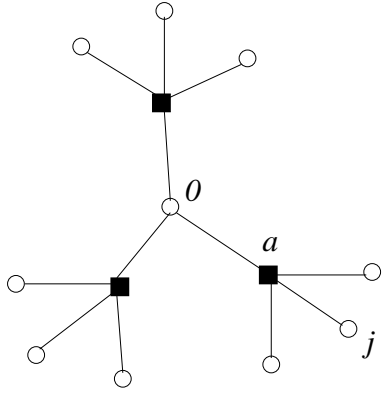}
\end{center}
\caption{Radius $1$ neighborhood of a typical site in the $i$
in the Tanner graph.}
\label{fig:Neigh1}
\end{figure}
We are left with the task of computing $p_{i,r}(x_i = \0t |\uy_{i,r})$.
As a warmup exercise, let us consider the case $r=1$. 
Without loss of generality, we set $i=0$. Because of the 
remark made in the previous Section, the subgraph $\B(0,1)$ is, with
high probability, a tree and must look like the graph in 
Fig.~\ref{fig:Neigh1}. Using the notations introduced in this figure, 
and neglecting normalization constants (which can be computed at 
the very end), we have
\begin{eqnarray}
p_{0,1}(x_0  |\uy_{0,1}) \propto \sum_{\{x_{j}\}}
Q(y_0|x_0)
\prod_{a\in\partial 0}
\prod_{j\in\partial a\backslash 0}Q(y_j|x_j)\,
\prod_{a\in\partial 0}\ind(x_{0}\oplus x_{ \partial a\backslash 0 }=0)\, .
\end{eqnarray}
Here we used the notation $\partial i$ ($\partial a$) to denote the set of
check nodes (respectively variable nodes) adjacent to variable node $i$
(resp. to check node $a$). Moreover, for $A=\{i_1,\dots,i_k\}$, we
wrote $x_A=x_{i_1}\oplus\cdots\oplus x_{i_k}$.
Rearranging the various summations, we get the expression
\begin{eqnarray}
p_{0,1}(x_0  |\uy_{0,1}) \propto Q(y_0|x_0) 
 \prod_{a\in\partial 0} \sum_{x_{j},\; j\in \partial a\backslash 0}
\ind(x_{0}\oplus x_{ \partial a\backslash 0 }=0)
\prod_{j\in\partial a\backslash 0}Q(y_j|x_j)\,\, , 
\end{eqnarray}
which is much simpler to evaluate due to its recursive structure. 
In order to stress this point, we can write the above formula as
\begin{eqnarray}
p_{0,1}(x_0  |\uy_{0,1})& \propto  & Q(y_0|x_0)
\prod_{a\in\partial 0} \mh_{a\to 0}(x_0) \, ,\\
\mh_{a\to 0}(x_0) & \propto & \sum_{x_{j},\; j\in \partial a\backslash 0}
\ind(x_{0}\oplus x_{ \partial a\backslash 0 }=0)
\prod_{j\in\partial a\backslash 0}\mu_{j\to a}(x_j)\, ,\\
\mu_{j\to a}(x_j) & \propto & Q(y_i|x_i)\, .
\end{eqnarray}
The quantities $\{\mu_{j\to a}(x_j)\}$,  $\{\mh_{j\to a}(x_j)\}$
are normalized distributions associated with the directed edges
of $G$. They are referred to as {\bf beliefs} or, more
generally, {\bf messages}. Notice that, in the computation of
$p_{0,1}(x_0  |\uy_{0,1})$ only messages along edges in
$\B(0,1)$, directed toward the site $0$, were relevant.

\begin{figure}
\begin{center}
\includegraphics[width=9cm]{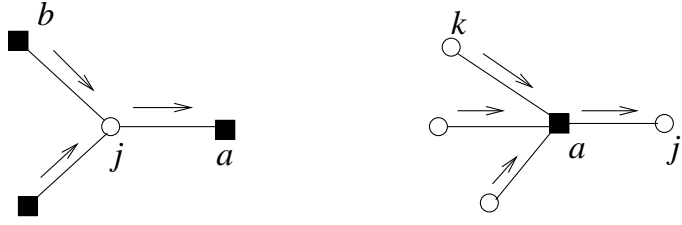}
\end{center}
\caption{Graphical representation of the belief propagation equations.}
\label{fig:BPEquations}
\end{figure}
In the last form, the computation of $p_{i,r}(x_i|\uy_{i,r})$ is easily
generalized to any finite $r$. We first notice that that
$\B(i,r)$ is a tree with high probability. Therefore, we can condition 
on this event without much harm. Then we associate messages
to the directed edges of $\B(i,r)$ (only messages directed towards $i$
are necessary). Messages are computed according to the rules
\begin{eqnarray}
\mu_{j\to a}(x_j)& \propto  & Q(y_j|x_j)
\prod_{b\in\partial j\backslash a} \mh_{b\to j}(x_j) \, ,
\label{eq:BPeq1}\\
\mh_{a\to j}(x_j) & \propto & \sum_{x_{k},\; k\in \partial a\backslash j}
\ind(x_{j}\oplus x_{ \partial a\backslash j }=0)
\prod_{k\in\partial a\backslash j}\mu_{k\to a}(x_j)\, ,
\label{eq:BPeq2}
\end{eqnarray}
with boundary condition $\mh_{b\to j}(x_j) = 1/2$ for all $b$'s outside
$\B(i,r)$. 
These equations are represented graphically in
Fig.~\ref{fig:BPEquations}. Finally, the desired marginal distribution
is obtained as 
\begin{eqnarray}
p_{i,r}(x_i  |\uy_{i,r})& \propto  & Q(y_i|x_i)
\prod_{a\in\partial i} \mh_{a\to i}(x_i)\, .\label{eq:MarginalFinal}
\end{eqnarray}

Let us now forget for a moment our objective of proving 
an upper bound on the conditional entropy $H(X_i|\uY)$. 
The intuitive picture is that, as $r$ increases, the marginal 
$p_{i,r}(x_i|\uy_{i,r})$ incorporates information coming form a 
larger number of received symbols and  becomes a more accurate approximation
of $p(x_i|\uy)$. 
Ideally, optimal decoding of the received message would require the
computation of $p(x_i|\uy)$, for which no efficient algorithm
is known. In particular, the expected number of incorrect bits
is minimized by the rule $\hat{x}(\uy) = \argmax_{x_i}\, p(x_i|\uy)$.
We can however hope that nearly optimal performances can be
obtained through the
rule
\begin{eqnarray}
\hat{x}_{i,r}(\uy) = \argmax_{x_i}
\, p_{i,r}(x_i|\uy_{i,r})\, .\label{BP:decoding}
\end{eqnarray}
Furthermore, a moment of thought shows that the recursive procedure
described above can be implemented in parallel for all the variables
$i\in [N]$. We just need to initialize $\mh_{b\to j}(x_j) = 1/2$
for all the check-to-variable messages, and then iterate the
update equations (\ref{eq:BPeq1}), (\ref{eq:BPeq2}) at all
nodes in $G$, exactly $r$ times.
For any fixed $r$, this requires just $\Theta(N)$ operations which is 
probably the smallest computational effort one can hope for.

Finally, although Eqs.~(\ref{eq:BPeq1}), (\ref{eq:BPeq2}) only 
allow to compute $p_{i,r}(x_i|\uy_{i,r})$ as far as 
$\B(i,r)$ is a tree, the algorithm is well defined for any value
of $r$. One can hope to improve the performances by taking larger 
values of $r$.
%
%
\subsection{A parenthesis}

The algorithm we `discovered' in the previous Section is in
fact well known under the name of {\bf belief propagation} (BP)
and is widely adopted for decoding codes on graphs.
This is in turn an example of a wider class of algorithms which are 
particularly adapted to problems defined on sparse graphs, and are
called (in a self-explanatory way) {\bf message passing algorithms}.
We refer to Sec.~\ref{sec:Historical} for some history and bibliography.

Physicists will quickly recognize that Eqs.~(\ref{eq:BPeq1}), (\ref{eq:BPeq2})
are just the equations for Bethe-Peierls approximation in the model
at hand \cite{SaadTAP,Yedidia}. 
Unlike the original Bethe equations, because of the quenched disorder,
the solutions of these equations depend on the particular sample,
and are not `translation invariant'.  These two features make
Eqs.~(\ref{eq:BPeq1}), (\ref{eq:BPeq2}) analogous to Thouless, 
Anderson, Palmer (TAP) equations for mean field spin glasses.
In fact Eqs.~(\ref{eq:BPeq1}), (\ref{eq:BPeq2}) are indeed the correct
generalization of TAP equations for diluted 
models~\footnote{In the theory of mean field disordered spin models, one
speaks of diluted models whenever the number of interaction terms
($M$ in the present case) scales as the size of the system.}.
Of course many of the classical issues in the context of the TAP 
(such as the existence of multiple solutions,
treated in the lectures of Parisi at this School) approach have a direct 
algorithmic interpretation here.

Belief propagation was introduced in the previous paragraph
as an algorithm for approximately computing the marginals of
the probability distribution
\begin{eqnarray}
p(\ux |\uy) \equiv \frac{1}{Z(\uy)}\;
\prod_{j\in [N]}Q(y_j|x_j)\, 
\prod_{a\in [M]}\ind(x_{i^a_1}\oplus\cdots\oplus x_{i^a_k}=0)\, ,
\end{eqnarray}
which is `naturally'  associated to the graph $G$.
It is however clear that the functions $Q(y_j|x_j)$ and 
$\ind(x_{i^a_1}\oplus\cdots\oplus x_{i^a_k}=0)$ do not have 
anything special. They could be replaced by any set of 
{\bf compatibility functions} $\psi_i(x_i)$, $\psi_a(x_{i^a_1},\dots , 
x_{i^a_k})$. Equations (\ref{eq:BPeq1}), (\ref{eq:BPeq2})
are immediately generalized. BP can therefore be regarded as a general 
inference algorithm for probabilistic graphical models.

The success of message passing decoding has stimulated several
new applications of this strategy: let us mention a few of them.
M\'ezard, Parisi and Zecchina \cite{MarcGiorgioRiccardo}
introduced `survey propagation', an algorithm
which is meant to generalize BP to the case in which the underlying 
probability distribution decomposes in an exponential number of 
pure states (replica symmetry breaking, RSB). In agreement with the prediction 
of RSB for many families of random combinatorial optimization problems,
survey propagation proved extremely effective in this context.
 
One interesting feature of BP is its decentralized nature. One
can imagine that computation is performed locally at variable and
check nodes. This is particularly interesting for applications
in which a large number of elements with moderate computational
power must perform some computation collectively (as is the case in 
sensor networks). Van Roy and Moallemi  \cite{VanRoy}
proposed a `consensus propagation'
algorithm for accomplishing some of these tasks.

It is sometimes the case that inference must be carried out in a situation 
where a well established probabilistic model is not available.
One possibility in this case is to perform `parametric inference'
(roughly speaking, some parameters of the model are left free).
Sturmfels proposed a `polytope propagation' algorithm for these cases
\cite{Sturmfels,SturmfelsAll}.
%
%
\section{Density evolution a.k.a. distributional recursive equations}

Evaluating the upper bound (\ref{eq:UpperBoundEXIT})
on the conditional entropy
described in the previous Section, is essentially the same 
as analyzing the BP decoding algorithm defined by 
Eqs.~(\ref{eq:BPeq1}) and (\ref{eq:BPeq2}). In order to accomplish this 
task, it is convenient to notice that distributions over a binary variable 
can be parametrized by a single real number. It is customary to choose this
parameters to be the {\bf log-likelihood ratios}
(the present definition differs by a factor $1/2$ from the traditional one):
\begin{eqnarray}
v_{i\to a}\equiv \frac{1}{2}\,\log\frac{\mu_{i\to a}(\0t)}{\mu_{i\to a}(\1t)}
\, ,\;\;\;\;\;\;\;\;\;
\vh_{a\to i}\equiv \frac{1}{2}\,\log\frac{\mh_{a\to i}(\0t)}
{\mu_{a\to i}(\1t)}\, .
\end{eqnarray}
We further define 
$h_i \equiv \frac{1}{2}\,\log\frac{Q(y_i|\0t)}{Q(y_i|\1t)}$.
In terms of these quantities, Eqs.~(\ref{eq:BPeq1}) and (\ref{eq:BPeq2})
read
\begin{eqnarray}
v^{(r+1)}_{j\to a} = h_j+\sum_{b\in\partial j\backslash a}\vh^{(r)}_{b\to j}\, ,
\;\;\;\;\;\;\;\;
\vh^{(r)}_{a\to j} = \atanh\left[\prod_{k\in\partial a\backslash j}\tanh 
v^{(r)}_{k\to a}\right] \, .
\end{eqnarray}
Notice that we added an index $r\in\{0,1,2,\dots\}$ that can be interpreted in 
two equivalent ways. On the one hand, the message $v^{r}_{j\to a}$
conveys information on the bit $x_j$ coming from a (`directed')
neighborhood of radius $r$. On the other $r$ indicates the number of
iterations in the BP algorithm. As for the messages, we can encode the 
conditional distribution $p_{i,r}(x_i|\uy_{i,r})$ through the 
single number $u_i^{(r)} \equiv \frac{1}{2}
\log \frac{p_{i,r}(\0t|\uy_{i,r})}{p_{i,r}(\1t|\uy_{i,r})}$. Equation 
(\ref{eq:MarginalFinal}) then reads
\begin{eqnarray}
u_i^{(r+1)} =h_i+\sum_{b\in\partial i}\vh^{(r)}_{b\to i}\, .
\label{eq:EffectiveField}
\end{eqnarray}

Assume now that the graph $G$ is distributed accordingly to the
$ \G_N(\Lambda,\Rho)$ ensemble and that $i\to a$ is a uniformly random 
(directed) edge in this graph. It is intuitively clear 
(and can be proved as well) that, for any giver $r$
$v^{(r)}_{i\to a}$ will converge in distribution to some well defined random 
variable $v^{(r)}$. This can be defined recursively by setting $v^{(0)}\ed h$
and, for any $r\ge 0$  
\begin{eqnarray}
v^{(r+1)} \ed h+\sum_{b=1}^{l-1}\vh^{(r)}_{b}\, ,
\;\;\;\;\;\;\;\;
\vh^{(r)} \ed \atanh\left[\prod_{j=1}^{k-1}\tanh 
v^{(r)}_{j}\right] \, ,\label{eq:DensityEvolution}
\end{eqnarray}
where $\{\vh^{(r)}_b\}$ are i.i.d. random variables distributed as
$\vh^{(r)}$, and $\{v^{(r)}_j\}$ are i.i.d. random variables distributed as
$v^{(r)}$. Furthermore $l$ and $k$ are integer random variable with 
distributions, respectively $\lambda_l$, $\rho_k$ 
depending on the code ensemble:
\begin{eqnarray}
\lambda_l = \frac{l\Lambda_l}{\sum_{l'}l'\Lambda_{l'}}\, ,\;\;\;\;\;\;\;
\;\;\;\;\;\;\;
\rho_k = \frac{k\Rho_k}{\sum_{k'}k'\Rho_{k'}}\, .
\end{eqnarray}
In other words $v^{(r)}$ is the message at the root of a random tree
whose offspring distributions are given by $\lambda_l$, $\rho_k$.
This is exactly the asymptotic distribution of the tree inside the ball 
$\B(i,r)$.
The recursions (\ref{eq:DensityEvolution}) are known in coding theory  as 
{\bf density evolution} equations. They are indeed the same (in the 
present context) as Aldous' 
recursive distributional equations \cite{AldousSteele},
or replica symmetric equations in spin glass theory.
From Eq.~(\ref{eq:EffectiveField}) one easily deduces that 
$u^{(r)}_i$ converges in distribution to $u^{(r)}$ defined by
$u^{(r+1)} \ed h+\sum_{b=1}^l\vh^{(r)}_{b}$ with $l$ distributed according 
to $\Lambda_l$. 

At this point we can use Eq.~(\ref{eq:UpperBoundEXIT})
to derive a bound on the bitwise entropy $h_{\rm b}$. Denote
by $h(u)$ the entropy of a binary variable whose log-likelihood ratio
is $u$. Explicitly 
\begin{eqnarray}
h(u) = -(1+e^{-2u})^{-1}\log(1+e^{-2u})^{-1}-
 (1+e^{2u})^{-1}\log(1+e^{2u})^{-1}\, . \label{eq:EXIT}
\end{eqnarray}
Then we have, for any $r$,
\begin{eqnarray}
\lim_{N\to\infty}\E_{\cC}\, h_{\rm b} \le \E\, h(u^{(r)})\, ,
\label{eq:BoundFinal}
\end{eqnarray}
where we emphasized that the expectation on the left hand side has to be 
taken with respect to the code. 

It is easy to show that the right hand side of the above inequality 
is non-increasing with $r$. It is therefore important to study its 
asymptotic behavior.
As $r\to \infty$, the random variables $v^{(r)}$ converge to a
limit $v^{(\infty)}$ which  depends on the code ensemble as well as on the
channel transition probabilities $Q(y|x)$. Usually one is interested in a 
continuous family of channels indexed by a noise parameter $p$ as for
the $\BSC(p)$, indexed in such a way that the channel `worsen' as
$p$ increases (this notion can be made precise and is called
{\bf physical degradation}). 
For `good' code ensembles, the following scenario holds.
For small enough $p$,
$v^{(\infty)}=+\infty$ with probability one. Above some critical value 
$p_{\sBP}$,
  $v^{(\infty)}\le 0$ with non-zero probability.
It can be shown that no intermediate case is possible.
In the first case BP is able to recover the transmitted codeword
(apart, eventually, from a vanishingly small fraction of bits),
and the bound (\ref{eq:BoundFinal}) yields $\E_{\cC}\, h_{\rm b}\to 0$.
In the second the upper bound remains strictly positive in the $r\to\infty$
limit.

A particularly simple channel model, allowing to work out in detail 
the behavior of density evolution, is the binary erasure channel
BEC$(p)$. In this case the channel output can take the values
$\{\0t,\1t,\?\}$. The transition probabilities are
$Q(\0t|\0t)=Q(\1t|\1t) =1-p$ and $Q(\?|\0t)=Q(\?|\1t) =p$. In other words,
each input is erased independently with probability $p$,
and transmitted incorrupted otherwise. We shall further assume, for the
sake of simplicity, that the random variables $l$ and $k$ in 
Eq.~(\ref{eq:DensityEvolution}) are indeed deterministic. In the other
words all variable nodes (parity check nodes) in the Tanner graph
have degree $l$ (degree $k$).
\begin{figure}
\centering
\setlength{\unitlength}{1bp}
\begin{picture}(400,70)
\put(0,20){\includegraphics[scale=0.5]{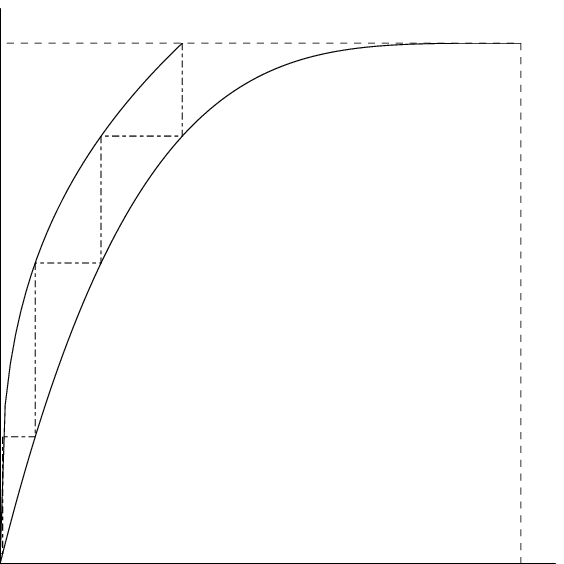}} 
\put(100,20){\includegraphics[scale=0.5]{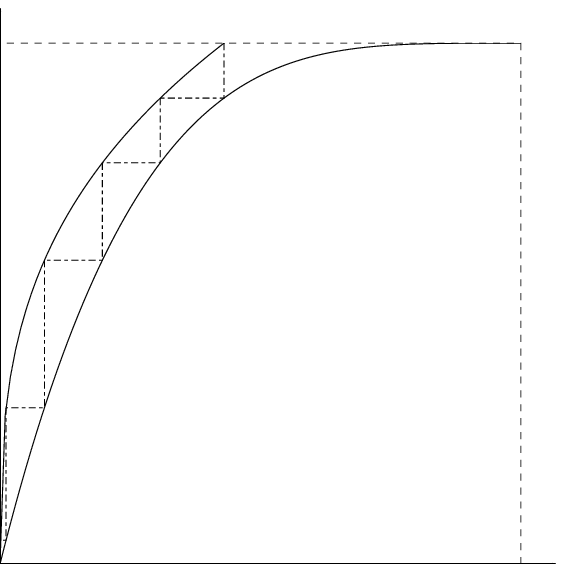}}
\put(200,20){\includegraphics[scale=0.5]{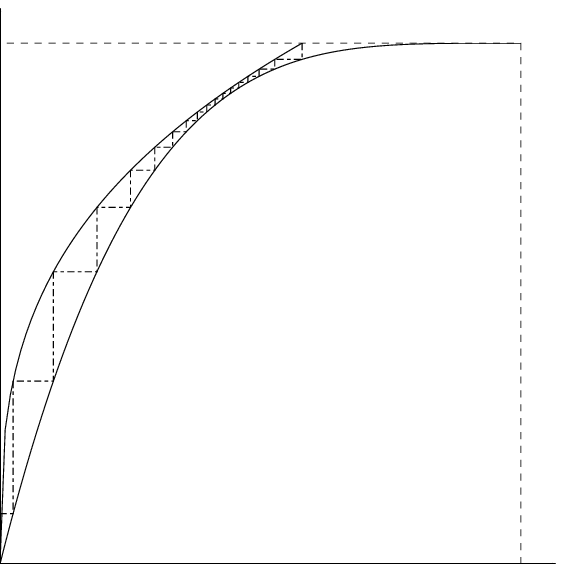}}
\put(300,20){\includegraphics[scale=0.5]{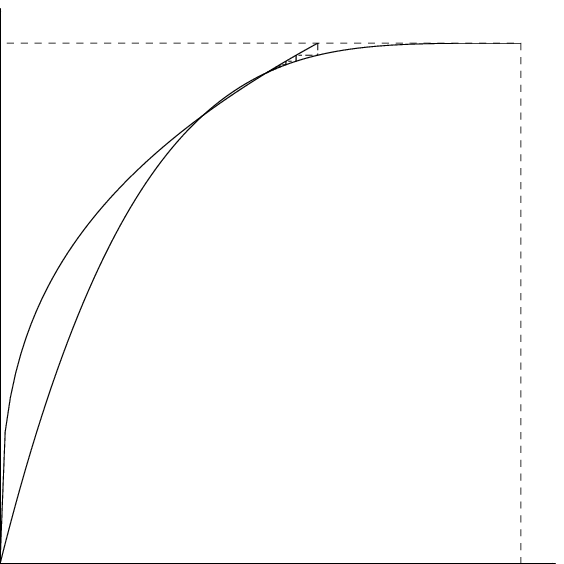}}
\put(-5,15){\makebox(0,0){\footnotesize{$0$}}}
\put(-5,95){\makebox(0,0){\footnotesize{$1$}}}
\put(80,15){\makebox(0,0){\footnotesize{$1$}}}
\put(40,5){\makebox(0,0){\footnotesize{$p=0.35< p_{\sBP}$}}}
\put(100,0)
{
\put(-5,15){\makebox(0,0){\footnotesize{$0$}}}
\put(-5,95){\makebox(0,0){\footnotesize{$1$}}}
\put(80,15){\makebox(0,0){\footnotesize{$1$}}}
\put(40,5){\makebox(0,0){\footnotesize{$p=0.43<p_{\sBP}$}}}
}
\put(200,0)
{
\put(-5,15){\makebox(0,0){\footnotesize{$0$}}}
\put(-5,95){\makebox(0,0){\footnotesize{$1$}}}
\put(80,15){\makebox(0,0){\footnotesize{$1$}}}
\put(40,5){\makebox(0,0){\footnotesize{$p=0.58<p_{\sBP}$}}}
}
\put(300,0)
{
\put(-5,15){\makebox(0,0){\footnotesize{$0$}}}
\put(-5,95){\makebox(0,0){\footnotesize{$1$}}}
\put(80,15){\makebox(0,0){\footnotesize{$1$}}}
\put(40,5){\makebox(0,0){\footnotesize{$p=0.61>p_{\sBP}$}}}
}
\end{picture}
\caption{Graphical representation of density evolution 
for the binary erasure channel, cf. Eq.~(\ref{BECRecursion}).}
\label{fig:iterativetrajectory}
\end{figure}
It is not hard to realize, under the assumption
that the all zero-codeword has been transmitted, that 
in this case $v^{(r)}$ takes values $0$ or $+\infty$. If 
we denote by $z_r$ the probability that $v^{(r)}=0$, the density evolution
equations  (\ref{eq:DensityEvolution}) become simply
\begin{eqnarray}
z_{r+1} = p\,\left(1-(1-z_r)^{k-1}\right)^{l-1}\, .\label{BECRecursion}
\end{eqnarray}
The functions $f_p(z) \equiv (z/p)^{1/(l-1)}$
and $g(z)\equiv 1-(1-z)^{k-1}$ are plotted in
Fig.~\ref{fig:iterativetrajectory} 
for $l=4$, $k=5$ and a few values of $p$ approaching $p_{\sBP}$. 
The recursion (\ref{BECRecursion}) can be described
as `bouncing back and forth' between the curves
$f_p(z)$ and $g(z)$. A little calculus shows
that $z_r\to 0$ if $p< p_{\sBP}$ while $z_r\to z_*(p)>0$
for  $p\ge p_{\sBP}$, where $p_{\sBP}\approx 0.6001110$ in the case 
$l=4$, $k=5$. A simple exercise for the reader  is to
work out the upper bound on the r.h.s. of Eq.~(\ref{eq:BoundFinal})
for this case and studying it as a function of $p$.

%
%
\section{The area theorem and some general questions}

Let us finally notice that general information theory considerations 
imply that $H(\uX|\uY)\le \sum_i H(X_i|\uY)$. As a consequence  
the total entropy per bit  $H(\uX|\uY)/N$ vanishes as well for 
$p<p_{\sBP}$. However this inequality greatly overestimates 
$H(\uX|\uY)$: bits entering in the same parity check are, for instance,
highly correlated. How can a better estimate be obtained?

The bound in Eq.~(\ref{eq:BoundFinal}) can be expressed by saying
that the actual entropy is strictly smaller than the one 
`seen' by BP. Does it become strictly positive for $p>p_{\sBP}$ because 
of the sup-optimality of belief propagation or because 
$H(\uX|\uY)/N$ is genuinely positive?

More in general, below $p_{\sBP}$ BP is essentially optimal. What happens
above? A way to state more precisely this question
consists in defining the distortion
\begin{eqnarray}
D_{\sBP,r} \equiv \frac{1}{N}\sum_{i=1}^N \sum_{x_i\in\{\0t,\1t\}}
\left| p(x_i |\uy)- p_{i,r}(x_i |\uy)\right|\, ,
\end{eqnarray}
which measures the distance between the BP marginals and the actual ones.
Below $p_{\sBP}$, $D_{\sBP,r}\to 0$ as $r\to\infty$. What happens above?

It turns out that all of these questions are strictly related.
We shall briefly sketch an answer to the first one and refer to the literature
for the others. However, it is worth discussing why they are challenging,
considering in particular the last one (which somehow implies the others).
Both $p(x_i|\uy)$ and $p_{i,r}(x_i |\uy)$ can be regarded
as marginals of some distribution on the variables associated to the 
tree $\B(i,r)$. While, in the second case, this distribution 
has the form (\ref{eq:TreeDistr}), in the first one some complicated 
(and correlated) boundary condition must be added in order to keep into account
the effect of the code outside $\B(i,r)$. Life would be easy 
if the distribution of $x_i$ were asymptotically decorrelated from the boundary
condition as $r\to\infty$, for {\em any boundary condition}. 
In mathematical physics terms, the infinite tree (obtained by taking 
$r\to\infty$ limit after $N\to\infty$) supports a {\em unique Gibbs measure}
\cite{Tatikonda}. In this case $p(x_i|\uy)$ and $p_{i,r}(x_i |\uy)$
simply correspond to two different boundary conditions and must coincide
as $r\to\infty$. Unhappily, it is easy to convince oneself that 
this is never the case for good codes! In this case no degree 0 or 1
variables exists and a fixed boundary condition always determines uniquely
$x_i$ (and more than one such condition is admitted).
 
As promised above, we conclude by explaining how to obtain 
a better estimate of the conditional entropy $H(\uX|\uY)$.
It turns out that this also provides a tool to tackle the other questions
above, but we will not explain how.
Denote by $w_i = 
\frac{1}{2}\log\frac{p(x_i=\0t|\uy_{\sim i})}{p(x_i=\1t|\uy_{\sim i})}$
the log-likelihood ratio which keeps into account all the
information pertaining bits $x_j$, with $j$ different from $i$,
and let $w^{(r)}_i$ be the corresponding $r$-iterations 
BP estimates. Finally, let $w^{(r)}$ be the weak limit of $w^{(r)}_i$
(this is given by density evolution, in terms of $v^{(r-1)}$).  
We introduce the so-called {\bf GEXIT} function $g(w)$.
For the channel $\BSC(p)$ this reads
\begin{eqnarray}
g(w) = \log_2\left\{1+\frac{1-p}{p}\, e^{-2w}\right\}-\log_2\left\{1
+\frac{p}{1-p}\, e^{-2w}\right\}\, .
\end{eqnarray}
And a general definition can be found in \cite{Life}. It turns out that
$\E g(w^{(r)})$ is a decreasing function of $r$ (in this respect, it is
similar to the entropy kernel $h(u)$, cf. Eq.~(\ref{eq:EXIT})). 
Remarkably, the following {\bf area theorem} holds
\begin{eqnarray}
H(\uX|\uY(p_1))-H(\uX|\uY(p_0))=\sum_{i=1}^N\int_{p_0}^{p_1} g(w_i)\,\de p\, ,
\end{eqnarray}
where $\uY(p_0)$,  $\uY(p_1)$ denotes the output upon transmitting through
channels with noise levels $p_0$ and $p_1$. Estimating the $w_i$'s
through their BP version, fixing $p_0=1/2$ (we stick, for the
sake of simplicity to the $\BSC(p)$ case)
and noticing that $H(\uX|\uY(1/2))=NR$, one gets
\begin{eqnarray}
H(\uX|\uY(p))/N\ge R-\int_{p}^{1/2}\!\!\!\! \E\, g(w^{(r)})\;\de p\, .
\end{eqnarray}
The bound obtained by taking $r\to\infty$ on the r.h.s. is expected to be 
asymptotically (as $N\to\infty$) exact for a large variety of code ensembles.
%
%
\section{Historical and bibliographical note}
\label{sec:Historical}

Information theory and the very idea of random code ensembles were 
first formulated by Claude Shannon in \cite{Shannon}. Random 
code constructions were never taken seriously from a practical point of view 
until the invention of turbo codes by Claude Berrou and Alain Glavieux in 
1993 \cite{Berrou}. This motivated a large amount of theoretical
work on sparse graph codes and iterative decoding methods.
An important step was the `re-discovery' of low density parity check codes,
which were invented in 1963 by Robert Gallager \cite{Gallager}
but soon forgotten afterwards. For an introduction to the subject and a
more comprehensive list of references see \cite{RichardsonUrbanke}
as well as the upcoming book \cite{RichardsonUrbankeBook}. See also
\cite{Factor} for a more general introduction to belief propagation 
with particular attention to coding applications.

The conditional entropy (or mutual information) for this systems was initially 
computed using non-rigorous statistical mechanics methods
\cite{Saad2,GallagerAM,NostroDynamical} 
using a correspondence first found by Nicolas Sourlas
\cite{Sourlas1}. These results were later proved to provide a 
lower  bound using
Guerra's interpolation technique \cite{Tight}, cf. also Francesco Guerra's
lectures at this School. Finally, an independent (rigorous)
approach based on the area theorem was developed in 
\cite{Life,MaxwellGeneral} and matching upper bounds were proved in 
particular cases in \cite{MaxwellBEC}.
%
%

\end{document}